\documentclass[pra,twocolumn,aps,amsmath,showkeys,showpacs,floatfix,nofootinbib]
{revtex4-1}

\usepackage{graphicx}
\usepackage{dcolumn}
\usepackage{bm}
\usepackage{color}

\newcommand{\ket}[1]{\left|#1\right\rangle} 
\newcommand{\bra}[1]{\left\langle#1\right|} 

\def\vec#1{{\boldsymbol{#1}}}  

\begin{document}


\title{Quantum chaos in $SU_3$ models with trapped ions}

\author{ Tobias Gra\ss$^1$, Bruno Juli\'a-D\'iaz$^{1,2}$, Marek Ku{\'s}$^3$, 
Maciej Lewenstein$^{1,4}$}

\affiliation{$^1$ICFO-Institut de Ci\`encies Fot\`oniques, Parc Mediterrani
de la Tecnologia, 08860 Barcelona, Spain}
\affiliation{$^2$  Departament d'Estructura i Constituents de la Mat\`{e}ria,
Universitat de Barcelona, 08028 Barcelona, Spain}
\affiliation{$^3$ Center for Theoretical Physics, Polish Academy of Sciences Al.
Lotnikow 32/46, 02-668 Warszawa, Poland}
\affiliation{$^4$ ICREA - Instituci\'o Catalana de Recerca i Estudis Avan\c
cats, 08010 Barcelona, Spain}

\begin{abstract}

A scheme to generate long-range spin-spin interactions 
between three-level ions in a chain is presented, providing a 
feasible experimental route to the rich physics of well-known 
$SU_3$ models. In particular, we demonstrate different signatures 
of quantum chaos which can be controlled and observed in  
experiments with trapped ions.

\end{abstract}

\maketitle


One of the current trends in quantum physics is the quest for 
controllable quantum many-body systems which can be used as 
quantum simulators~\cite{mlbook, orbital}. In particular, 
there is a growing interest in simulating spin and quantum 
magnetism.  In recent years, the focus is moving from $SU_2$ 
spins towards $SU_N$~\cite{Gorshkov, zollersun}, which can be 
realized in earth alkalines, or mixed spin spaces~\cite{schmidt-kaler}. Here we
show an implementation of $SU_3$ physics with trapped ions which are known to
provide a large degree of control from the experimental point of view.

An important feature of quantum simulators based on ions is 
the possibility of  studying long-range interactions, which are 
notoriously difficult to simulate 
classically~\cite{PhysRevLett.109.267203,luca}. The implementation 
is based on spin dependent forces on the ions~\cite{porras04}, 
which have been experimentally achieved 
recently~\cite{schaetz-natphys,monroe-spinspin,britton2012engineered,blattross}.
These interactions lead 
to new phases, like exotic forms of superfluidity~\cite{capo}, 
supersolids~\cite{pollet}, quantum crystals, and devil's 
staircase~\cite{PhysRevB.80.174519,hauke-devil}.

We concentrate on an important aspect present in $SU_3$ 
models: quantum chaos~\cite{haake-book,gnutzmann}. Quantum 
chaos, opposed to classical chaos which can be defined by 
exponentially fast growing distance of phase space trajectories, 
was strongly driven by the understanding of the spectral 
properties of quantum many-body systems~\cite{PhysRevLett.52.1}. 
The large degree of control offered by experiments with 
ultracold atomic gases has triggered a vast number of 
experiments to look for different signatures of 
chaos~\cite{Raizen:2011}. Prominent examples are the 
observation of dynamical tunneling phenomena~\cite{Hensinger, Steck13072001}, 
and more recently, the implementation of the kicked-top 
Hamiltonian on a single atom experiment~\cite{Chaudhury}. 
Recent proposals look for signatures of chaotic behavior 
in spin-orbit coupled condensates~\cite{jonas} or in kicked 
Bose-Hubbard dimers~\cite{christine}. 

In this letter, we demonstrate that the extremely long-range 
character of interactions between ions can be used to mimic 
shell models which are paradigmatic of quantum 
chaos~\cite{haake-book,gnutzmann,PhysRevLett.79.4790}. 
We calculate  experimentally controllable signatures of chaos, 
and estimate the fidelity of the proposed simulation in the
Supplementary Material.

\vspace{0.3cm}
{\bf Spin-spin interactions of ions:}
The main ingredient required to achieve a strong and controllable spin-spin interaction between trapped ions is the implementation 
of a state-dependent force on the ions. In an early proposal by Mintert
and Wunderlich~\cite{mintert01}, such force is induced 
by a magnetic field gradient. More flexible proposals provide a force by a
Raman coupling of two spin states, which can give rise to phonon excitations.
While the proposal by Porras and Cirac~\cite{porras04} uses standing waves in
all spatial direction, and thereby allows for up to three independent spin-spin
interactions, experimental implementations have provided Ising-type spin-spin
interactions using a Raman coupling due to a pair of propagating waves with a
wave vector difference transverse to the ion alignment
\cite{schaetz-natphys,monroe-spinspin}. By choosing the
frequency difference between the two lasers close to a vibrational sideband
transition, one can enhance and control the phonon excitations, and
thus
the Ising coupling constants. In particular, single vibrational modes can be
selected.

\begin {figure}
\includegraphics[width=0.47\textwidth]{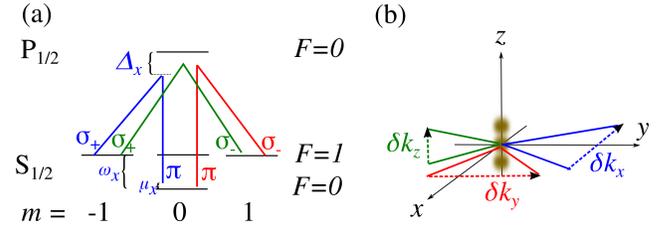}
\caption{(Color online) \label{Fig1} 
(a) Level structure of the ions: Three levels are pairwise coupled by
far-detuned Raman lasers, with a beat note $\mu_\alpha$ providing a
spin-dependent force (see text). (b) Arrangement of the Raman lasers, with
polarizations given in (a) and wave vector difference $\delta k_\alpha$ along
the axes.}
\end {figure}

Here we generalize this scheme to three couplings in a system of three-level
ions, which allows for an implementation of SU(3) spin models.
All spin states, for instance represented by three hyperfine states
$\ket{1} \equiv \ket{F=0,m=0}$, $\ket{2} \equiv \ket{F=1,m=-1}$, and $\ket{3}
\equiv \ket{F=1,m=1}$ in the $S_{1/2}$ manifold of Yb$^+$, are coupled as
depicted in Fig.~\ref{Fig1}(a), via off-resonant transitions to an excited state
$\ket{\rm e} \equiv \ket{F=0}$ in the $P_{1/2}$ manifold. The lasers are
arranged as shown in Fig.~\ref{Fig1}(b), such that
each Raman pair has a wave vector difference $\delta \vec{k}_{\alpha}$ along a
different spatial direction $\alpha=x,y,z$. Interference between different
couplings can be avoided by choosing different detunings $\Delta_{\alpha}$ from
$\ket{\rm e}$.

Each coupling can be equipped with beat notes $\omega_{\alpha} \pm
\mu_{\alpha}$, where $\mu_{\alpha}$ describes the detuning from the
atomic transition frequency $\omega_{\alpha}$, given by the hyperfine splitting
and/or the Zeeman splitting. After adiabatic elimination of the excited state
$\ket{\rm e}$ and a rotating wave approximation, the Hamiltonian of
each coupling reads \cite{monroe-spinspin}:
\begin{align}
\label{halpha}
 h_{\alpha}(t) = \hbar \sum_i \Omega_{\alpha}^{(i)} \sin(\mu_{\alpha}t) \delta \vec{k}_{\alpha} \cdot
\vec{x}^{(i)} \tau_{\alpha}^{(i)}.
\end{align}
Here, $\Omega_{\alpha}^{(i)}$ is the two-photon Rabi frequency of the Raman
transition $\alpha$ at the position of ion $i$.
The spin-flip operators are defined as $\tau_x \equiv |1><2| + \rm{H.c}$,
$\tau_y \equiv |1><3| + \rm{H.c}$, and $\tau_z \equiv |2><3| + \rm{H.c.}$.
Next, we rewrite the position operator in terms of normal coordinates,
$\delta \vec{k}_{\alpha} \cdot \vec{x}^{(i)} \equiv \sum_{m}  \eta_{m \alpha}^{(i)} (a_{m\alpha} e^{-i \omega_{m\alpha} t} + a_{m\alpha}^{\dagger} e^{i \omega_{m\alpha} t})$, where vibrational modes $m$ along direction $\alpha$
 are summed, with $\omega_{m\alpha}$, $a_{m\alpha}$, and $a_{m\alpha}^{\dagger}$
the corresponding frequency, annihilation, and creation operator. We have
introduced the Lamb-Dicke parameter $\eta_{m\alpha}^{(i)}$ characterizing the
strength of the spin-phonon coupling. For the validity of Hamiltonian
(\ref{halpha}), it is necessary to have $\eta_{m\alpha}^{(i)} \ll 1$. It is
explicitly given by $\eta_{m\alpha}^{(i)}=\sqrt{ \frac{\hbar}{2M\omega_{m\alpha}
}} {\cal M}^{\alpha}_{m,i}$. The $N \times N$ matrix ${\cal M}^{\alpha}_{m,i}$
are the normal modes in $\alpha$ direction, found by ${\cal M}^{\alpha}_{m,i}
{\cal K}^{\alpha}_{m m'}{\cal M}^{\alpha}_{m',i'} = \omega_{m\alpha}^2
\delta_{i,i'}$. The kernel $\cal K$ contains the Coulomb repulsion and the
external trapping of frequency $\omega_{\alpha}$ along each direction.  Assuming
linearly arranged and equidistant equilibrium positions, it reads:
\begin{eqnarray}
{\cal K}^{\alpha}_{m,m'} 
&= \left\{ \begin{array}{ll} \omega^2_{\alpha} - c_{\alpha} \sum_{m'' (\neq m)}
\frac{1}{|m-m''|^3}, \hspace{0.5cm} & m=m'
\\ + c_{\alpha}
\frac{1}{|m-m'|^3}, & m \neq m' \end{array}
\right.
\end{eqnarray}
where $c_{x,y} = 1$, $c_z = - 2$. We have chosen ``ionic'' units, in 
which besides the electric constant $1/(4\pi \epsilon_0)$ also 
the ion mass $M$, the ion charge $q$, and the equilibrium distance 
$d$ of neighboring ions in the chain are set to unity. Frequencies 
are then given in units of $\omega_0 \equiv q/(d \sqrt{4\pi\epsilon_0 d M})$.

It has been shown in Ref. \cite{monroe-spinspin} that the time evolution under
the Hamiltonian of Eq. (\ref{halpha}) is (to second order in the exponent) given
by
\begin{align}
U_\alpha(t,0)=\exp \left[ \sum_{i} \varphi_\alpha^{(i)}(t) \tau_{\alpha}^{(i)} -
\sum_{ij} \xi_{\alpha}^{(i,j)}(t) \tau_\alpha^{(i)} \tau_\alpha^{(j)} \right],
\end{align}
where $\varphi_{\alpha}^{(i)} = \sum_m( c_{m\alpha}^{(i)}(t)
a_{m\alpha}^{\dagger} - \rm{H.c.})$ contains a residual spin-phonon coupling,
while the second term describes a spin-spin coupling. Both functions $
c_{m\alpha}^{(i)}(t)$ and $\xi_{\alpha}^{(i,j)}(t)$ consist of oscillatory
terms (with frequencies $\mu_\alpha$, $\mu_\alpha \pm \omega_{m\alpha}$),
which are suppressed by at least one power of
$\eta_{m\alpha}^{(i)}\Omega_{\alpha}^{(i)}/|\mu_\alpha - \omega_{m\alpha}|
\ll 1$ for sufficiently large detuning from the sideband. In this limit, the
dominant contribution to the time evolution stems from a single term in
$\xi_{\alpha}^{(i,j)}(t)$ which is linear in $t$, and thus increases constantly.
Thus, we can set $c_{m\alpha}^{(i)} \approx 0$, and $\xi_{\alpha}^{(i,j)}(t)
\approx i J_{ij}^\alpha t$, with
\begin{align}
\label{Jij}
 J^{(i,j)}_\alpha = \Omega_{\alpha}^{(i)}\Omega_{\alpha}^{(j)}   \sum_m \frac{
\eta_{m\alpha}^{(i)}\eta_{m\alpha}^{(j)} \omega_{m\alpha}
}{\mu_\alpha^2 - \omega_{m\alpha}^2}.
\end{align}
The time evolution is thus identical to the one of a spin model with spin-spin
coupling $J^{(i,j)}_\alpha$.

In the presence of more than one couplings, since $[h_\alpha,h_\beta] \neq 0$,
the time evolution is not simply the product of all $U_\alpha$, but
consists of additional terms. Up to second order in the
Magnus expansion \cite{Blanes}, it is equal
to $U \simeq (\prod_\alpha
U_\alpha) (\prod_{\alpha \neq \beta}
U_{\alpha\beta})$ with
$U_{\alpha\beta}=\exp\left\{\sum_{ij}
\chi_{\alpha\beta}^{(i,j)}(t)[\tau_{\alpha}^{(i)},\tau_{\beta}^{(j)}]\right\} $.
The functions $\chi_{\alpha\beta}^{(i,j)}$ are given by the integral
\begin{align}
 \chi_{\alpha\beta}^{(i,j)} =& \sum_{m,n} \int_{0}^{t} \mathrm{d}t_1
\int_{0}^{t_1} \mathrm{d}t_2 \ \sin{\mu_\alpha t_1}\sin{\mu_\beta
t_2} 
\ \times \\ \nonumber &
(a_{m\alpha} e^{i \omega_{m\alpha}t_1}+ {\rm H.c.})(
a_{m\beta} e^{i\omega_{m\beta}t_2} +{\rm H.c.}).
\end{align}
For $\mu_\alpha = \mu_\beta$ and $\omega_{m\alpha} = \omega_{n\beta}$, this
function is similar to $\xi_{\alpha}^{(i,j)}(t)$ with small oscillating terms,
and one dominant term linear in $t$. However, if couplings $\alpha$ and $\beta$
have different beat notes and/or the mode frequencies are different (for
instance by anisotropic transverse trapping), the linear term disappears, and
this contribution can be neglected.

In addition to the spin-spin coupling, a magnetic field
term can be generated by a resonant carrier transition
\cite{schaetz-natphys,kim2010}, leading to the effective spin Hamiltonian
\begin{align}
 H_{\rm spin} = \sum_{\alpha} \left[ \sum_{i=1}^N B^{(i)}_\alpha
\sigma_{z}^{(i)}+
\sum_{i\leq j}^N
J^{(i,j)}_{\alpha} \tau^{(i)}_{\alpha} \tau^{(j)}_{\alpha} \right],
\label{ham}
\end{align}
with $\sigma_z^{(i)} \equiv \ket{1}\bra{1}^{(i)} - \ket{3}\bra{3}^{(i)}$. The
couplings $J^{(i,j)}_\alpha$ given by Eq.~(\ref{Jij}), can be tuned by the beat
notes $\mu_\alpha$. In particular, by choosing $\mu_\alpha$ sufficiently close
to the center-of-mass (COM) mode in each direction $J^{(i,j)}_\alpha$ will have
only weak dependence on the ion positions $i$ and $j$. Most generally, we assume
an inhomogeneous magnetic field $B_{\alpha}^{(i)}$.

\vspace{0.3cm}
{\bf Effective $SU_3$ Shell Model:}
In the limit where $J^{(i,j)}_{\alpha}=J_{\alpha}={\rm constant}<0$, 
it is convenient to define spin-flip operators $S_{\sigma\sigma'} = \sum_{l=1}^N
\ket{\sigma}\bra{\sigma'}^{(l)}$, 
acting equally on \textit{all} spins. Since we have $S_{11}+S_{22}+S_{33}=N$, 
the $S_{\sigma\sigma'}$ provide eight independent operators spanning the $SU_3$ 
algebra. For simplicity, we set $J_\alpha=J$ and the magnetic field 
homogeneous. Defining a symmetrized spin operator 
$\tilde S_{\sigma\sigma'} \equiv (S_{\sigma\sigma'} +
S_{\sigma'\sigma})/\sqrt{2}$, we may re-write 
the spin Hamiltonian of Eq.~(\ref{ham}) as an ideal model 
Hamiltonian in terms of these $SU_3$ operators:
\begin{align}
 H_{\rm ideal} = \frac{B}{\sqrt{2}} (\tilde S_{11}- \tilde S_{33}) + J
\sum_{\sigma<\sigma'} \tilde S_{\sigma\sigma'} \tilde S_{\sigma\sigma'}.
\label{hspin}
\end{align}
Besides the replacement $S_{\sigma\sigma'} \rightarrow \tilde
S_{\sigma\sigma'}$, this 
Hamiltonian is identical to the three-level Lipkin-Meshkov-Glick 
(LMG) Hamiltonian~\cite{lipkin}, a model, where particles can occupy three
different shells with 
single-particle energies $-B,0,B$. Two-body interactions of 
particles in the same shell lead to pair-tunneling into the 
other shells. The LMG Hamiltonian has applications in nuclear 
physics, and its three-level version is particularly appealing 
as a not fully integrable spin model in the context of quantum 
chaos~\cite{meredith,gnutzmann}.

Our Hamiltonian~(\ref{hspin}) differs from the LMG Hamitonian only by an
additional interaction $\sum_{\sigma \neq \sigma'}
S_{\sigma\sigma'}S_{\sigma'\sigma}$. Due to its $SU_3$
symmetry, and the fact that this additional term is a Casimir operator of 
$SU_3$~\cite{gnutzmann}, the Hamiltonian~(\ref{hspin}) and the
LMG Hamiltonian are fully equivalent: Having a block-diagonal structure
with respect to different representations of $SU_3$, in each block the
additional interaction simply reduces to a constant.

Apart from particle exchange symmetry, the LMG model has a 
second symmetry~\cite{meredith}: As particles can change 
the spin state only pairwise, the occupation numbers of 
each spin state, $\langle S_{11}\rangle$, $\langle S_{22}
\rangle$, and $\langle S_{33} \rangle$, can only change by 
two, and thus are fixed to either even ($e$) or odd ($o$) values. 
This gives rise to four signature classes, $eee$, $oeo$, $ooe$, 
$eoo$ for $N$ even, or $ooo$, $eeo$, $eoe$, $oee$ for $N$ odd.
This signature class symmetry is also present in Eq.~(\ref{ham}), 
that is, in the model with a space-dependent coupling $J_{\alpha}^{(i,j)}$ 
given by Eq.~(\ref{Jij}). On the other hand, the spin exchange 
symmetry is lost. Still there is invariance under parity, as 
$J^{(i,j)} = J^{(N-i)(N-j)}$, due to the parity invariance of 
${\cal K}$. For a numerical diagonalization of this Hamiltonian, 
it is convenient to construct the eigenbasis of parity and 
signature class. While the Fock states are already signature 
eigenstates, a combination of at most two Fock states also 
yields a parity eigenstate.

\begin {figure}[t]
\includegraphics[width=0.49\textwidth]{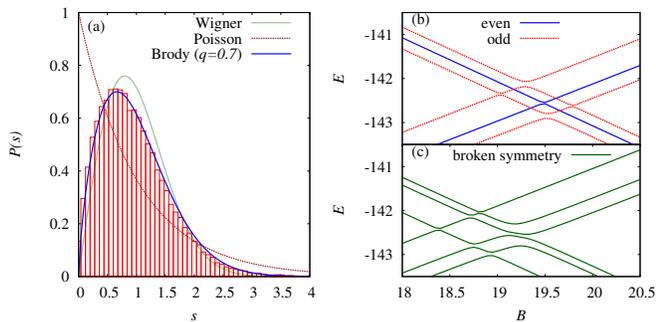}
\caption{(Color online) \label{Fig3} (a) Level spacing distribution for $N=10$
after unfolding the spectrum separately in each symmetry block of the
Hamiltonian.
(b) Avoided crossing of energy levels with equal parity symmetry, exemplified
for $N=4$ in the \textit{eoe} signature class upon tuning the magnetic field
strength $B$. (c) The same as in (b), but in the presence of a small additional
magnetic field gradient $\delta_B=0.2$ breaking the parity symmetry. This leads
to
avoided crossings of all energy levels.}
\end {figure}

\vspace{0.3cm}
{\bf Quantum chaos in the LMG model:}
In the classical limit of the three-level LMG model, its 
phase-space can be divided into regions of chaotic and 
regular motion~\cite{meredith}. Accordingly, also the 
quantum model shows signatures of both chaotic and regular 
behavior. While in chaotic quantum systems the spectrum 
features level repulsion, regular behavior is related to 
level clustering. These features are nicely displayed by 
the unfolded distribution~\cite{haake-book} of the 
level spacings $s$ in the spectrum. A Poisson distribution, 
$P(s)=e^{-s}$, indicates level clustering, while chaotic 
Hamiltonians with time-reversal invariance follow a 
Wigner distribution, $P(s)=(\pi/2)s \exp[-\pi s^2/4]$. 
In Ref.~\cite{meredith}, it has been shown for the LMG 
model that one part of the spectrum is spaced according 
to the Poisson distribution, while another part follows a Wigner 
spacing.

This results in a level spacing distribution as shown in Fig.~\ref{Fig3}(a) 
for $N=10$ and a magnetic field $B=\langle J_\alpha^{(i,j)} \rangle/2$. 
We have used the realistic Hamiltonian (\ref{ham}), with a trap frequency
$\omega_\alpha=0.1 \omega_0 / \delta_\alpha$ at a relative detuning from the COM
mode, $\delta_\alpha = (\omega_{{\rm COM} \alpha} - \mu_\alpha)/\omega_{{\rm
COM} \alpha} > 0$. As is shown in the Supplementary 
Material, this Hamiltonian reproduces with high fidelity the physics of the 
ideal model, Eq.~(\ref{hspin}). We have unfolded the spectrum 
separately in each symmetry block of the Hamiltonian (that is for 
fixed parity and signature class). The level spacing distribution is 
found to be broader than the Wigner 
distribution, and has its maximum shifted towards smaller spacings. 
This suggests to consider the Brody distribution $P_q(s)$ which 
interpolates between the Wigner ($q=1$) and the Poisson 
distribution 
$(q=0)$~\cite{brody,wunner}:
\begin{align}
P_q(s) = \alpha(q+1) s^q  \exp[-\alpha s^{q+1}],
\end{align}
with $\alpha=[\Gamma((q+2)/(q+1))]^{q+1}$. The value of $q$
provides a measure of the degree of chaoticity in the system. 
As shown in Fig.~\ref{Fig3}(a), our distribution is well represented 
by $q=0.7$.

The behavior expressed by these statistics can be illustrated 
by representing the evolution of a few energy levels when one 
parameter of the Hamiltonian is changed, e.g. the magnetic field 
strength $B$. In each symmetry block, we find both level crossings 
and avoided level crossings, as already expected from the level 
spacing distribution. In Fig.~\ref{Fig3}(b), we illustrate, for 
$N=4$, a part of the spectrum where all level crossings belonging 
to states of the same symmetry (parity) are avoided. Of course, 
the crossings between states of different parity are not avoided. 
In Fig.~\ref{Fig3}(c), we then show the same part of the spectrum 
in the presence of an additional small magnetic field gradient 
$\delta_B=0.2$, that is for an inhomogeneous magnetic field 
$B_{\rm inhom}(x)= B + \delta_B x$. 
This contribution breaks the parity symmetry, turning the 
previously symmetry allowed level crossings, Fig.~\ref{Fig3}(b), 
into avoided ones.

\begin {figure}[t]
\includegraphics[width=0.49\textwidth]{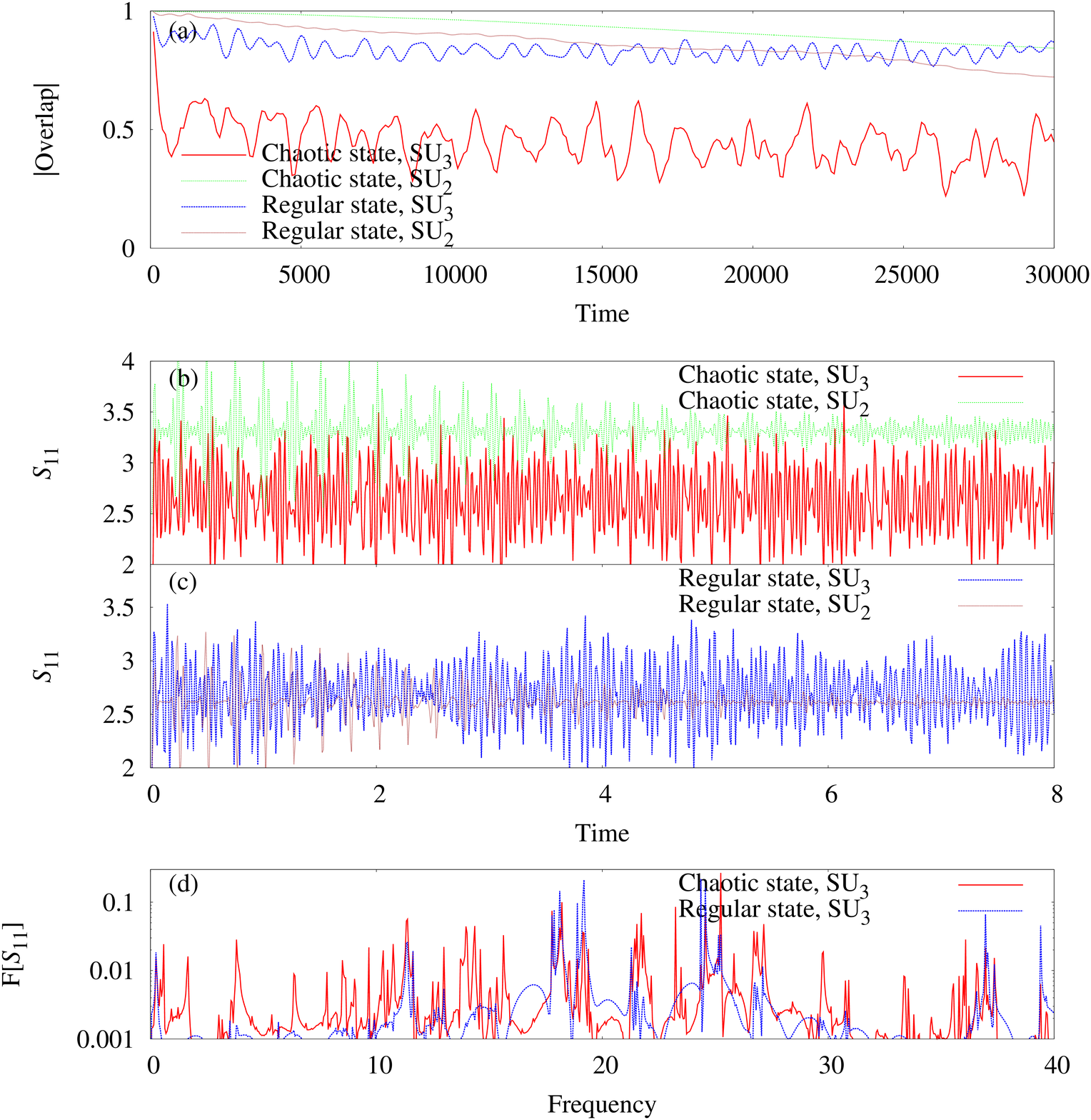}
\caption{(Color online) \label{Fig4} (a) We evolve one 
chaotic state, $z_1=-0.10+0.61 i$ and $z_2=-0.83+0.26i$, 
and one regular  state, $z_1=-1.06+0.26i$ and $z_2=-1.04+0.33i$, 
with $N=8$ particles in the 
Hamiltonian $H_{\rm spin}(B,J_x,J_y,J_z)$ (\ref{ham}) for $B=0.5$ and 
$B=0.505$. For each initial state, we plot the overlap between 
the evolved states for the two $B$s as a function of time. The 
curves labeled with $SU_3$ are obtained by choosing all 
interactions $J_{\alpha}$ to be transmitted by equally strong
forces $K_\alpha$, while the $SU_2$ curves are obtained for 
$J_x=J_y=0$. (b,c) For the same states as in (a) and with 
$B=0.5$, we plot the occupation number $\langle S_{11} \rangle$ 
as a function of time. (d) The Fourier transform of the $SU_3$ 
curves in (b,c).}
\end {figure}

\vspace{0.3cm}
{\bf Experimental detection of quantum chaos:}
The signatures of quantum chaos discussed so far are hard 
to measure in a system of trapped ions. More easily, quantum chaos can
be detected by preparing the system initially in a coherent 
quantum state, and then observing the subsequent time evolution 
of this state~\cite{Hensinger,Steck13072001,Chaudhury}. Unitarity 
of quantum evolution prevents a definition of quantum chaos 
mirroring the usual one in classical systems: exponential 
sensitivity to initial conditions. Instead, for quantum-chaotic 
motion it is argued that a relevant signature is provided by 
high sensitivity of the time evolution onto slight changes 
in the Hamiltonian parameters~\cite{haake-book}.

In order to relate our study of the quantum dynamics to its classical 
limit, the initial states will be $SU_3$ coherent spin states defined 
as $\ket{z_1, z_2} \equiv {\cal N} \exp[z_1 S_{31} + z_2 S_{21}] \ket{0}$, 
where $\ket{0}$ denotes a Fock state which is fully spin-polarized in the lower 
spin component, $\ket{0} \equiv \ket{1 1 \cdots 1}$, and ${\cal N}$ normalizes 
the state. The complex parameters $z_1, z_2$ define the classical 
state in terms of four canonical variables, $q_1,q_2,p_1,p_2$. The classical 
Hamiltonian is then given by~\cite{meredith}:
\begin{align}
 H_{\rm class}(q_1,q_2,p_1,p_2) = \lim_{N\rightarrow \infty}\bra{z_1,z_2} H_{\rm
ideal}/N \ket{z_1,z_2},
\end{align}
We have performed the classical time evolution using a Runge-Kutta method.
The coherent states with small average energy are mostly found to have
regular behavior, while states of intermediate energy behave rather
chaotically. 

We will now search for signatures of chaos in the quantum time
evolution, driven by the Hamiltonian $H_{\rm spin}$. We again choose
$\omega_\alpha=0.1 \omega_0/\delta_\alpha$, and consider two initial states: 
one in a classical regular region, 
and the second one in a classical chaotic region. As shown in 
Fig.~\ref{Fig4}(a), a minimal change of $1 \%$ in the parameter 
$B$ has little effect on the quantum time evolution of the 
regular state compared to its effect on the evolution of the 
chaotic state. This indicates that even for a system of 8 ions, 
far from the classical limit, we observe clear signatures of 
quantum chaotic behavior in correspondence with the expected 
behavior in the classical limit. For comparison, the figure 
also shows the time evolution of the same initial states for 
a Hamiltonian where by choosing $J_x=J_y=0$ one spin state 
has been dynamically frozen. In this way, the model reduces 
to an $SU_2$ LMG model~\cite{lipkin}, which is integrable, and 
accordingly shows no trace of quantum chaos.

While the overlaps shown in Fig.~\ref{Fig4}(a) are not directly 
accessible in experiments, signatures of chaotic behavior can 
also be found in the evolution of a spin component of the state.
In our scheme, the occupation of the level $\ket{m=0}$ can be measured with
high precision by resonantly exciting ions from this level, and observing the
subsequent fluorescence. For a system prepared in the regular state a regular
pattern is expected, while an erratic pattern is a signature of chaotic
motion~\cite{haake-book}. 
We exemplify this in Fig.~\ref{Fig4}(b,c), showing the time 
evolution of $\langle S_{11} \rangle$ for the regular and the 
chaotic state given above, evolved with $B=0.5$ in the full 
$SU_3$ Hamiltonian and in the reduced $SU_2$ Hamiltonian. The 
curves for $SU_2$ clearly show a regular pattern, while in the 
$SU_3$ case the differences between the chaotic and the
regular state are less obvious. We therefore perform a Fourier 
analysis of these curves after subtracting its average and 
normalizing the amplitude of the oscillation. In the Fourier 
spectrum, shown in Fig. \ref{Fig4}(d), the regular evolution is 
dominated by only a few peaks, while the spectrum of the chaotic
evolution is much more diversified.

\vspace{0.3cm}
{\bf Summary:} 
We have presented a scheme to realize $SU_3$ spin models with 
trapped ions. The spin-spin interaction parameter is tunable, allowing the
simulation of spin models with long-range interactions. In this paper,
we have focused on the situation where interactions are almost constant with
respect to the position of the ions. Then the systems describes a three-level
LMG model which interpolates between quasi-integrable and chaotic dynamics.
Chaos on the quantum level is usually characterized in terms of the
spectral statistics which demands high-precision measurements of many energy
levels, hardly possible in the settings similar to the one discussed in our
paper. Instead we propose a feasible scheme to detect quantum chaos in the
dynamics of the model. Our proposal thus provides a powerful experimental tool
to study the onset and signatures of chaos in quantum systems, based on
state-of-the-art techniques. 

\vspace{0.3cm}
{\bf Acknowledgements:}
This work has been supported by EU (NAMEQUAM, AQUTE),
ERC (QUAGATUA), Spanish MINCIN (FIS2008-00784 TOQATA),
Generalitat de Catalunya (2009-SGR1289), Polish NSC grant DEC-2011/02/A/ST1/00208, and 
Alexander von Humboldt Stiftung.
BJD is supported by the Ram\'on y Cajal program.


\end{document}